\documentclass[a4paper, aps, prd, reprint, twocolumn, amsmath,
  showpacs, superscriptaddress]{revtex4-1}
\usepackage{graphicx}
\def\aa{Astron.\ Astrophys.\ }
\def\apjl{Astrophys.\ J.\ Lett.\ }
\def\apjs{Astrophys.\ J.\ Suppl.\ }
\def\mnras{Mon.\ Not.\ R.\ Astron.\ Soc.\ }
\def\prsla{Proc.\ R.\ Soc.\ A\ }
\def\cqg{Classical\ Quantum\ Gravity\ }
\begin{document}
\title
{Unstable normal modes of low $T/W$ dynamical instabilities \break
in differentially rotating stars}
%
\author{Motoyuki Saijo}
\email[E-mail: ]{saijo@aoni.waseda.jp}
%
\affiliation
{Department of Science, Waseda University,
 Shinjuku, Tokyo 169-8050, Japan}
\altaffiliation[Present address: ]
{Department of Physics, Waseda University,
 Shinjuku, Tokyo 169-8555, Japan}
%
\author{Shin'ichirou Yoshida}
\email[E-mail: ]{yoshida@ea.c.u-tokyo.ac.jp}
%
\affiliation
{Graduate School of Arts and Science, The University of Tokyo,
 Meguro, Tokyo 153-8902, Japan}
%
\received{26 January 2016}
\revised{3 July 2016}
\accepted{2 October 2016}
%
\begin{abstract}
We investigate the nature of low $T/W$ dynamical instabilities in 
differentially rotating stars by means of linear perturbation.  Here, 
$T$ and $W$ represent rotational kinetic energy and the gravitational 
binding energy of the star.  This is the first attempt to investigate 
low $T/W$ dynamical instabilities as a complete set of the eigenvalue 
problem.  Our equilibrium configuration has ``constant'' specific 
angular momentum distribution, which potentially contains a singular 
solution in the perturbed enthalpy at corotation radius in linear 
perturbation.  We find the unstable normal modes of differentially 
rotating stars by solving the eigenvalue problem along the equatorial 
plane of the star, imposing the regularity condition on the center and 
the vanished enthalpy at the oscillating equatorial surface.  We find 
that the existing pulsation modes become unstable due to the existence 
of the corotation radius inside the star.  The feature of the unstable 
mode eigenfrequency and its eigenfunction in the linear analysis 
roughly agrees with that in three-dimensional hydrodynamical
simulations in Newtonian gravity.  Therefore, our normal mode analysis
in the equatorial motion proves valid to find the unstable equilibrium
stars efficiently.  Moreover, the nature of the eigenfunction that
oscillates between corotation and the surface radius for unstable
stars requires reinterpretation of the pulsation modes in
differentially rotating stars.
\end{abstract}
%
\pacs{04.25.Nx, 04.40.b, 97.10.Kc, 97.10.Sj}
\maketitle
%
\section{Introduction
\label{sec:intro}}
Low $T/W$ dynamical instability in differentially rotating stars was 
first discovered by numerical simulations \citep{PDD96, CNLB01, SKE02, 
  SKE03, SBS03}.  Here, $T$ and $W$ represent rotational kinetic
energy and gravitational binding energy of the star.  The instability
timescale is dynamical, and the spiral- and bar-type deformation is
found, but the strength of instability seems weaker than the standard
dynamical bar instability \citep{Chandra69, Tassoul, ST83}.  Since low 
$T/W$ dynamical instability takes place in a simple physical system of
self-gravitating differentially rotating objects, it is considered as
playing an essential role in many astrophysical scenarios of compact
objects.  Mergers of binary neutron stars may form differentially
rotating objects and trigger a spiral type of low $T/W$ dynamical
instabilities \citep{PEPS15, EPPS16}.  A spiral type of low $T/W$
dynamical instabilities may play an essential role in supernova 
explosion for efficient angular momentum transport after the core
bounce, e.g., Ref.~\citep{KTK14}.  The effect of fragmentation of the
star may be caused by the low $T/W$ dynamical instabilities
\citep{ZSHOSE06, ZSHOSE07, ROAHMS13}.  In all these scenarios, 
nonaxisymmetric deformation of the compact objects arises due to low
$T/W$ dynamical instabilities, and it generates gravitational waves
that are to be detected in the ground-based or the space-based
interferometer within next five years \citep{gw13}.

Among the variety of astrophysical applications of low $T/W$
dynamical instabilities, its mechanism is still unknown.  There are
several studies on corotation instabilities along the cylindrical star
\citep{Balbinski85} or along the accretion disk system with
Wentzel-Kramers-Brillouin (WKB) approximation \citep{PP87, TL08}.
There is an indication from the basic pulsation equations that the
corotation radius (the radius where the equilibrium fluid and pattern
rotate at the same angular speed) inside the star may play a role for
low $T/W$ dynamical instabilities \citep{WAJ05}.  For the rotating
stellar configuration, \citet{SY06} investigate in practice the
instabilities using the canonical angular momentum by both
hydrodynamical and perturbative approaches and find that the
corotation radius plays an essential role for angular momentum
transport (see also Ref.~\citep{OT06}).  Recently, properties of the
$f$ mode have been studied when corotation exists inside the stars in
the linearized Newtonian hydrodynamics \citep{PA15}.

In this paper, we focus on the nature of low $T/W$ dynamical
instabilities mainly from the normal mode analysis in the equatorial
motion of the star by the linear perturbation approach.  Our
particular concern is the investigation of the stability and its
nature of the system with eigenvalue analysis.  We introduce a
simplified one-dimensional eigenvalue problem in the equatorial
motion, with taking possible corotation singularity in perturbed
enthalpy into account (see also Ref.~\citep{TL08} for corotation
singularity in accretion disk system).  Although there is a study that 
existing $f$ mode is unstabilized when high degree of differential
rotation is taken into account \citep{SKE02, SKE03} or corotation
occurs \citep{PA15}, we never ascertain whether corotation singularity 
generates new types of pulsation modes in differentially rotating
stars without examining the complete set of eigenmode analysis.  Full
two-dimensional studies of linear eigenmodes of rapidly rotating stars 
have been done mainly in the uniformly rotating stars, e.g.,
Refs.~\citep{IL89, LRR06}.  The effect of differential rotation is
considered \citep{SKE02, SKE03, KE03, RMJSM09, ODR12}, but the
detailed examination of relation between corotation singularity and
the excitation of low $T/W$ dynamical instability remains beyond their
study.  Therefore, we simply focus on the equatorial motion, which can
easily handle corotation singularity.  We then investigate the nature
of the eigenfunction of the mode from the viewpoint of the existence
of corotation inside the stars.  Finally, comparison of our eigenmode 
analysis in the perturbative approach with those in three-dimensional
hydrodynamics clearly shows that our model contains sufficient valid
physics.  Throughout this paper, we use gravitational units with
$G=1$.

\section{Linear Perturbation
\label{sec:bequation}}

Axisymmetric equilibrium configuration of differentially rotating
stars in Newtonian gravity can be constructed numerically by using the 
technique established by Hachisu \citep{Hachisu86, SK08}.  We impose
perfect fluid with a polytropic equation of state
$p=\kappa\rho^{\Gamma}$, where $p$ is the pressure, $\rho$ is the rest
mass density, $\kappa$ is the constant, $\Gamma= 1 + 1/n$ is the
adiabatic exponent, $n$ is the polytropic index, and $j$-constant
rotation law for the angular velocity distribution is 
\begin{equation*}
\Omega = \frac{j_0}{d^2 + \varpi^2},
\end{equation*}
where $j_0$ is the constant, $d$ is the degree of differential
rotation, and $\varpi$ is the radial distance of the cylindrical 
coordinates.  Here, we focus on the low $T/W$ dynamically unstable
star, which is summarized in Table~\ref{tab:table1}.  Note that
models~I and II represent $m=2$ and $m=1$ dominant dynamically
unstable stars. 

\begin{table}[b]
\begin{center}
\caption{
Equilibrium configuration of differentially rotating stars}
\begin{ruledtabular}
\begin{tabular}{c c c c c}
Model & $n$ & $r_p / r_e$\footnotemark[1] & $\Omega_c /
\Omega_e$\footnotemark[2] & $T/W$
 \\
 \hline
I & 1 & 0.625 & 26.0 & $6.09 \times 10^{-2}$\\
II & 3 & 0.625 & 26.0 & $7.21 \times 10^{-2}$\\
\end{tabular}
\end{ruledtabular}
\label{tab:table1}
\footnotetext[1]{$r_p$: Polar surface radius; $r_e$: Equatorial
  surface radius}
\footnotetext[2]{$\Omega_c$: Central angular velocity; $\Omega_e$:
  Equatorial surface angular velocity}
\end{center}
\end{table}

We perturb the differentially rotating stars nonaxisymmetrically in
order to investigate the feature of low $T/W$ dynamical
instabilities.  The nonaxisymmetrically perturbed quantity $\delta q$
has a dependence of $\delta q = \delta q_m (\varpi, z) e^{-i\omega t +
  i m \varphi}$, where $z$ is the coordinate along the rotational axes
and $\varphi$ is the azimuthal coordinate.

We impose one assumption in which the equatorial motion of the
perturbed quantities alone is taken into account.  Our basic idea is
that the characteristic wave propagation due to rotation mainly lies
in the equatorial plane.  The pulsation equations of differentially
rotating stars in Newtonian gravity in the equatorial motion can be
written by using the perturbed continuity equation, perturbed Euler
equation and perturbed Poisson equation as (e.g., Ref.~\citep{TL08})
\begin{eqnarray}
&&\left[
\frac{d^2}{d\varpi^2} - 
\left( \frac{d}{d\varpi} \ln \frac{D}{\rho \varpi} \right)
  \frac{d}{d\varpi} 
  - \frac{2m\Omega}{\varpi \tilde{\omega}}
\left( \frac{d}{d\varpi} \ln \frac{\rho \Omega}{D} \right) 
\right. \nonumber \\
&&
\hspace{1cm}
\left.
- \frac{m^2}{\varpi^2} - \frac{D}{dp/d\rho}
\right]
\delta U_m (\varpi) = - \frac{D}{dp/d\rho} \delta\Phi_m (\varpi)
\label{eqn:Basic_dU}
,\\
&&\left[
\frac{d^2}{d\varpi^2} + 
\frac{1}{\varpi} \frac{d}{d\varpi}
 - \frac{m^2}{\varpi^2} + 4 \pi \rho \frac{d\rho}{dp} 
\right]
\delta \Phi_m (\varpi) \nonumber \\
&&
\hspace{1cm}
= 4\pi\rho \frac{d\rho}{dp} \delta U_m (\varpi),
\label{eqn:Basic_dPhi}
\end{eqnarray}
where $\delta U_m$ is the scalar potential $\delta U_m \equiv \delta
h_m + \delta \Phi_m$, $\delta h_m$ is the perturbed enthalpy, and
$\delta \Phi_m$ is the perturbed gravitational potential; $D =
\kappa^2 - \tilde{\omega}^2$, with $\kappa^2$ being $\varpi (d\Omega^2
/ d\varpi) + 4 \Omega^2$ and $\tilde{\omega} = \omega - m \Omega$.
Note that we discard the second-order $z$ derivative in $\delta U_m$
and $\delta \Phi_m$ (the first-order $z$ derivative in $\delta U_m$
and $\delta \Phi_m$ automatically disappears due to the equatorial
symmetry we imposed in the system) to derive Eqs.~(\ref{eqn:Basic_dU})
and (\ref{eqn:Basic_dPhi}).  The equations contain singular solution
around $\tilde{\omega} = 0$.  Expanding the equations around the
corotation radius (the radius $ r_{\rm cr}$ where $\omega = m\Omega$
in a pure real frequency), we can easily find the singular solution in
$\delta U_m$ as
\begin{eqnarray}
\delta U_m &=& A_{1m} x + A_{2m} 
  \bigl[ 1 + |\beta| x (-1 + 2 \gamma + \log |\beta|
\nonumber \\
&& \hspace{1cm}
   + \log |x|) + O(x^2)\bigr],\\
\beta &=&
- \frac{2 \Omega_{\rm cr}}{r_{\rm cr} d\Omega/d\varpi|_{\varpi=r_{\rm cr}}} 
\left. \frac{d}{d\varpi}
  \left( \ln \frac{\kappa^2}{\rho \Omega} \right)
\right|_{r=r_{\rm cr}} < 0,
\end{eqnarray}
where $x\equiv \varpi - r_{\rm cr}$, $A_{1m}$ and $A_{2m}$ are
constants, $\gamma$ is the Euler's constant, and $\Omega_{\rm cr}$ is
the angular velocity at corotation.  Since $\delta U_m$ contains the
term $x \log |x|$, $d\delta U_m / d\varpi$ contains singular behavior
at corotation.  To construct a solution across corotation in a pure
real frequency ($\Im[\omega]=0$), analytic continuation is necessary
on corotation.  Since $\delta \Phi_m$ is regular from the fact that
$\delta U_m$ is continuous on the corotation [see
Eq.~(\ref{eqn:Basic_dPhi})], only the singular behavior (discontinuity
in the first derivative of $\delta U_m$) is contained in $\delta
h_m$.

To check our assumption, we compare the results of linear perturbation
in the cylindrical model with those of hydrodynamics in
Sec.~\ref{sec:hydro}.

\begin{table}[b]
\begin{center}
\caption{
$m=1$ and $m=2$ normal modes of differentially rotating stars in the 
  cylindrical model}
\begin{ruledtabular}
\begin{tabular}{c c c c c c}
Model & $m$ & $N$\footnotemark[1]  & $\Re [\omega] / \Omega_c $ & $\Im
[\omega] / \Omega_c $ &
$r_{\rm cr} / r_e$\footnotemark[2]
\\
\hline
I & $1$ & 1 & 0.65275 & 0.00012 & 0.14587\\
I & $1$ & 2 & 1.15197 & 0.00000 & $\cdots$\\
\hline
I & $2$ & 0 & 0.29017 & 0.00855 & 0.48549\\
I & $2$ & 1 & 0.89791 & 0.00100 & 0.22158\\
I & $2$ & 2 & 1.39597 & 0.00039 & 0.13156\\
I & $2$ & 3 & 1.86074 & 0.00024 & 0.05471\\
I & $2$ & 4 & 2.31080 & 0.00000 & $\cdots$\\
\hline
\hline
II & $1$ & 0 & 0.52603 & 0.05126 & 0.18985\\
II & $1$ & 1 & 0.72759 & 0.00688 & 0.12238\\
II & $1$ & 2 & 0.97965 & 0.00011 & 0.02883\\
II & $1$ & 3 & 1.24636 & 0.00000 & $\cdots$\\
\hline
II & $2$ & 0 & 0.52656 & 0.00041 & 0.33456\\
II & $2$ & 1 & 0.71208 & 0.00147 & 0.26897\\
II & $2$ & 2 & 0.94196 & 0.00072 & 0.21197\\
II & $2$ & 3 & 1.17403 & 0.00039 & 0.16775\\
II & $2$ & 4 & 1.40432 & 0.00037 & 0.13026\\
II & $2$ & 5 & 1.63112 & 0.00038 & 0.09511\\
II & $2$ & 6 & 1.85820 & 0.00046 & 0.05525\\
II & $2$ & 7 & 2.08739 & 0.00000 & $\cdots$\\
\end{tabular}
\end{ruledtabular}
\label{tab:table2}
\footnotetext[1]{$N$: Node numbers between the corotation and surface 
  radius}
\footnotetext[2]{$r_{\rm cr}$: Corotation radius (the radius where
  $\Re[\omega] = m\Omega$)}
\end{center}
\end{table}

\begin{figure}
\centering
\includegraphics[keepaspectratio=true,width=8cm]{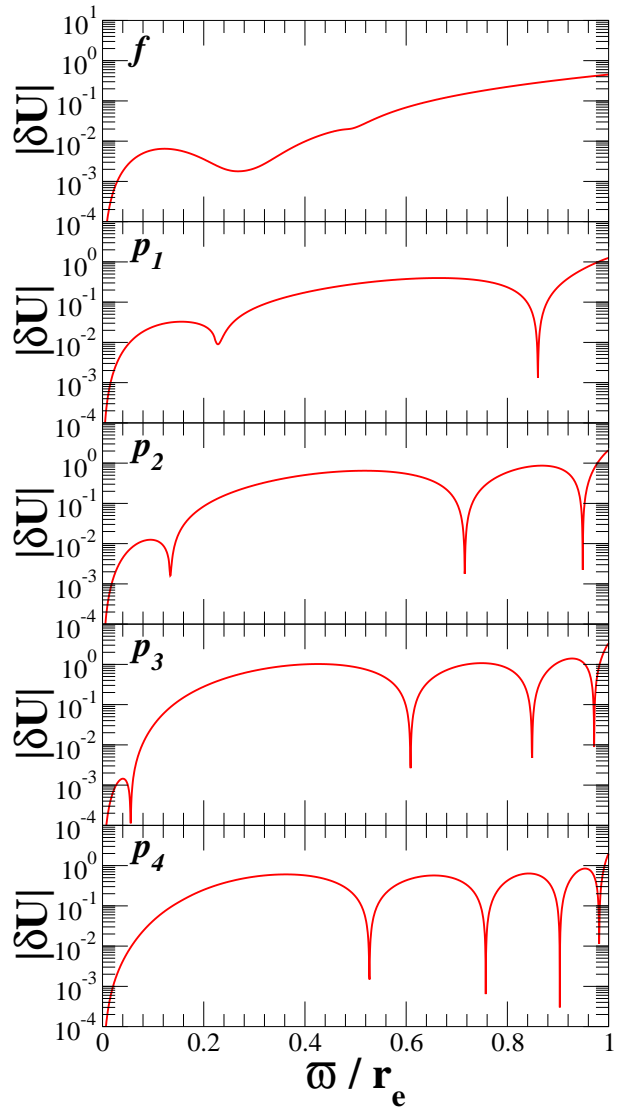}
\caption{
Eigenfunctions of the first five $m=2$ normal modes of differentially
rotating $n=1$ polytropic stars in the cylindrical model.  Labels $f$
and $p_N$ denote the $f$ mode and $p$ modes with node number $N$ in
Table~\ref{tab:table2}, respectively.  Note that node numbers counted
between corotation ($r_{\rm cr} / r_{e} = 0.486, 0.222, 0.132, 0.0564$
for $N=0$, $1$, $2$, $3$ normal modes) and the equatorial surface
radius are used to identify $f$ and $p$ modes.
\label{fig:figure1}
}
\end{figure}

\begin{figure}
\centering
\includegraphics[keepaspectratio=true,width=8cm]{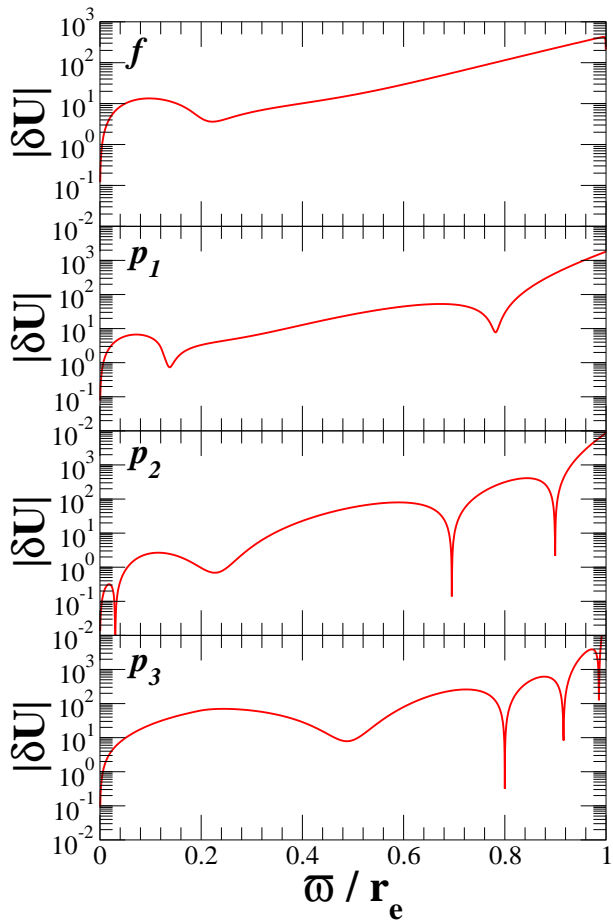}
\caption{
Same as Fig.~\ref{fig:figure1}, but for first three $m=1$ normal modes
of $n=3$ polytropic stars.  Note that the corotation radius for $N=0$,
$1$, and $2$ normal modes are $r_{\rm cr} / r_{e} = 0.190$, $0.122$,
and $0.0288$.
\label{fig:figure2}
}
\end{figure}

\section{Unstable normal modes
\label{sec:NM}}
We study the stability of the system by introducing eigenvalue
problem.  We impose regularity conditions at the center as 
\begin{equation*}
\delta U_m = C_{1m} \varpi^{|m|},
\hspace{5mm}
\delta \Phi_m = C_{2m} \varpi^{|m|},
\end{equation*}
where $C_{1m}$ and $C_{2m}$ are constants.  We set the boundary
condition for $\delta \Phi_m$ at infinity as the quantity is finite
($\delta \Phi_m \propto \varpi^{-|m|}$).  This is equivalent to
imposing a surface boundary condition as $\delta \Phi_m = C_{3m}
\varpi^{-|m|}$.  The constant $C_{3m}$ is described by an appropriate
combination of $C_{1m}$ and $C_{2m}$, which is determined from the
condition that $\delta \Phi_m$ and $d \delta \Phi_m / d\varpi$ are
continuous across the surface.  We also impose the surface boundary
condition for $\delta U_m$ as the enthalpy vanishes on the oscillating
surface of the stars.  Namely
\begin{equation}
\delta h_m + \xi^{j}_m \nabla_{j} h = 0,
\label{eqn:bc_cy}
\end{equation}
where $\xi^i_m$ is the Lagrangian displacement \citep{IL90},
$\nabla_{j}$ is the covariant derivative, and $h$ is the equilibrium
enthalpy.  Only 1 degree of freedom seems to remain in the system,
which represents the normalization factor.  We set $C_{1m}=1$ in our
computational code.

The axisymmetric equilibrium configuration of the differentially
rotating stars is computed in the two-dimensional cylindrical
coordinates \citep{SK08}.   Then we take the equilibrium quantities
$q$ ($\equiv p/\rho$) and $\Phi$ (gravitational potential) 3841 grid 
points along the equatorial plane from the center to the stellar
surface in order to integrate the pulsation equations.  We apply a
fourth-order Runge-Kutta method to integrate Eqs.~(\ref{eqn:Basic_dU})
and (\ref{eqn:Basic_dPhi}).

We search a complex frequency $\omega$ in the region of $\Re[\omega]
\in$ [$0$,$m+0.4$] $\Omega_c$ and $\Im[\omega] \in$ [$0$,$0.1$]
$\Omega_c$, with nondimensional resolution $\Delta \omega / \Omega_c = 
1 \times 10^{-5}$ for both real and imaginary frequencies.
Introducing a complex frequency $\omega$ avoids singular solution at
corotation ($\tilde{\omega} \not= 0$ for $\Im[\omega]\not= 0$ inside
the star).  We integrate Eqs.~(\ref{eqn:Basic_dU}) and
(\ref{eqn:Basic_dPhi}) from the center to the half-radius, and from
the surface with boundary conditions including Eq.~(\ref{eqn:bc_cy}) 
to the half-radius, and then match the solutions by computing the 
determinant, composed of two sets of solutions $\delta U_m$, $d\delta 
U_m / d\varpi$, $\delta \Phi_m$ and $d\delta \Phi_m / d\varpi$.  Only 
a successful choice of frequency can generate the solution from the 
center to the surface.  We determine the eigenfrequency when the 
absolute value of the determinant takes the minimum at a certain 
complex frequency by comparing with that at the neighbouring four 
frequencies in the complex frequency plane.  Since our approach cannot 
treat $\tilde{\omega}=0$ inside the star due to corotation
singularity, we simply discard the search region $\Re[\omega] \in$
[$0,m$] $\Omega_c$ and $\Im[\omega] \in$ [$0,0.2~\Delta \varpi / r_e$]
$\Omega_c$, where the nondimensional grid resolution $\Delta \varpi /
r_e \equiv 1/3840 \approx 2.6 \times 10^{-4}$.  Note that $\omega /
\Omega_c = 0.2~\Delta \varpi / r_e$ is a typical angular momentum
resolution around the surface ($\Delta \Omega /\Omega_c \gtrsim \Delta
\Omega_e / \Omega_c \approx 7.4 \times 10^{-2} \Delta \varpi / r_e$
for $\Omega_c / \Omega_e = 26$).  Here, we only focus on $m=1$ spiral
and $m=2$ bar modes.

We show the series of the complex eigenfunction in
Fig.~\ref{fig:figure1} for the $m=2$ mode of model~I and in
Fig.~\ref{fig:figure2} for the $m=1$ mode of model~II.  All the first
five nodes for $m=2$ and the first three for $m=1$ can be seen along
the radial direction in each eigenfunction.  For example, the
fundamental unstable mode ($N=0$ in Table~\ref{tab:table2}) has no
node (the eigenfunction does not cross zero), which represents the $f$
mode (e.g., Ref. \citep{Cox80}).  The other unstable modes in
Table~\ref{tab:table2} have $N$ nodes, i.e. the eigenfunctions cross 
zero $N$ times between the corotation and surface radius.  The
features can be interpreted to mean that these modes are $p$ modes 
(e.g., Ref. \citep{Cox80}).  The reason for counting the node numbers
between the corotation and surface radius is to reckon the strength of
the corotation barrier to each eigenfunction.  An unstable
eigenfunction has a sharp change at the corotation radius, since the
corotation singularity acts as a potential barrier for wave
propagation.  The sharp change in the eigenfunction can also be
interpreted as a phase gap in the eigenfunction, which arises through
analytic continuation on corotation.  We summarize our results of
complex eigenfrequencies in Table~\ref{tab:table2}.  We only find the
known types of modes ($f$ and $p$ modes) with every node numbers are
present in our analysis of the equatorial fluid motion, except for the
$f$ mode ($N=0$) in model~I.  Note that we discard the
``eigenfrequency'' which relates to a constant cylindrical
displacement \citep{Unno89}.  Also all eigenfrequencies which possess
corotation inside the star are found unstable.  The dominance of the
$m=1$ or $m=2$ mode depends on the stiffness of the equation of
state. In fact, the shortest timescale $\tau \equiv
(\Im[\omega])^{-1}$ (largest growth rate) from the normal mode
analysis for model~I is the case $m=2$ $f$  mode, while for model~II,
it is the case $m=1$ $f$ mode.  This feature has also been found in
numerical simulations \citep{SBS03, SY06}.

\section{Comparison with hydrodynamical simulation
\label{sec:hydro}}
We briefly introduce our three-dimensional hydrodynamical simulation 
in Newtonian gravity and compare the results with the linear analysis.
We compute the same differentially rotating equilibrium stars
summarized in Table~\ref{tab:table1} and impose nonaxisymmetric
perturbation in the rest mass density as 
\begin{equation*}
\rho = \sum_{k=1}^{4} \rho_{\rm eq}\left[ 1 + \delta_k
  \frac{\varpi}{r_e}(\cos k \varphi + \sin k \varphi) \right],
\end{equation*}
where we set $\delta_i = 5 \times 10^{-5}$ ($i=1, \cdots, 4$) for
evolution.  Note that $\rho_{\rm eq}$ is the equilibrium configuration
of the rest mass density, $x$ and $y$ are the components of Cartesian
coordinates, and the cylindrical radius $\varpi$ is $\varpi =
\sqrt{x^2+y^2}$.  We insert the approximate Harten-Lax-van Leer (HLL)
Riemann solver \citep{Toro09} with the same reconstruction method,
MC-limiter \citep{LeVeque98}, for hydrodynamics in our code
\citep{SK08}.  We have demonstrated the results of the wall shock
problem in our code, which are in full agreement with those of
one-dimensional analytical solution.  We monitor the diagnostics $M_1$
and $M_2$ (e.g., Ref.~\citep{SK08}), which are the $m=1$ and $m=2$
rest mass density-weighted average in the whole volume, and find that
both $M_1$ and $M_2$ grow exponentially in time for low $T/W$
dynamically unstable case.  Here we focus on the dominant $m$ mode for
each model (Table~\ref{tab:table1}).  In practice, the diagnostic
$M_2$ grows exponentially up to $t \approx 170~P_c$ for model~I and
$M_1$ to $t \approx 70~P_c$ for model~II, and saturates its amplitude
around $M_2 \approx 0.16$ for model~I and $M_1 \approx 0.006$ for
model~II.  Note that $P_c$ is the central rotation period of the
equilibrium star.  The diagnostic $M_1$ and $M_2$ clearly show that
the characteristic frequency is $\Re[\omega] = 0.387~\Omega_c$ for
model~I and $\Re[\omega] = 0.586~\Omega_c$ for model II, and the
growth timescale is $\tau \equiv (\Im[\omega_i ])^{-1} = 10.7~P_c$ for
model~I and $\tau \equiv (\Im[\omega_i ])^{-1} = 9.75~P_c$ for
model~II, both timescales being extracted from the first $60~P_c$.
Note that our timescales are typical for low $T/W$ dynamical
instability \citep{Muhlberger14}.  We extract the complex frequencies
by spectrum analysis \citep{SK08} for the real part and the fitting
formula of the dominant $m$ mode of $M_1$ and $M_2$ diagnostics for
the imaginary part.  Note that 161 grid points are covered along the
equatorial diameter of the star, with an equatorial radius twice as
large as the outer boundary for each coordinate direction.  We also
check the different grid resolution for accuracy in both
characteristic frequencies and growth timescales.  We find that the 
characteristic frequency changes 0.4\%--2\% in the relative error rate
and the growth timescale changes 20\%--40\%, the covered grid points
along the equatorial diameter of the star varying between 121 and 161,
fixing the ratio between the stellar radius and the outer boundary.
 
\begin{figure}
\centering
\includegraphics[keepaspectratio=true,width=8cm]{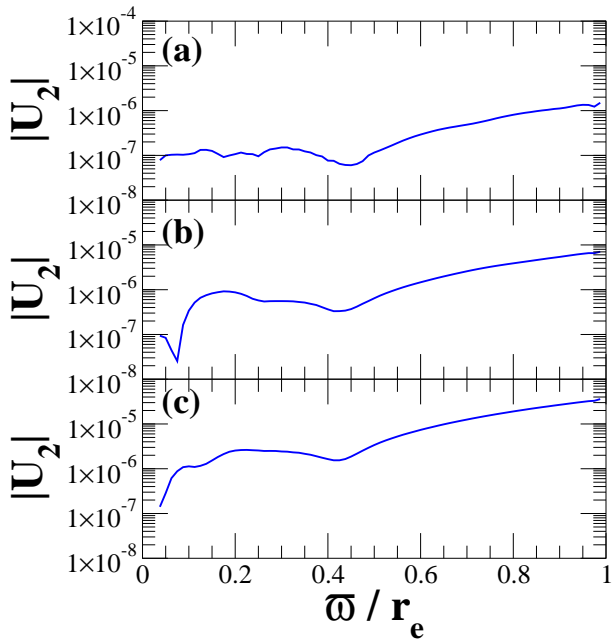}
\caption{
The $m=2$ scalar potential density diagnostics in the equatorial
plane.  Note that $U_2$ represents the $m=2$ scalar potential density
weighted average in the equatorial plane.  The labels (a), (b), and
(c) denote the evolution time $t= 59.7~P_c$, $74.6~P_c$, and
$89.5~P_c$, respectively.  We find features similar to the cylindrical
model (Fig.~\ref{fig:figure1}) in which no nodes are present.
\label{fig:figure3}
}
\end{figure}

\begin{figure}
\centering
\includegraphics[keepaspectratio=true,width=8cm]{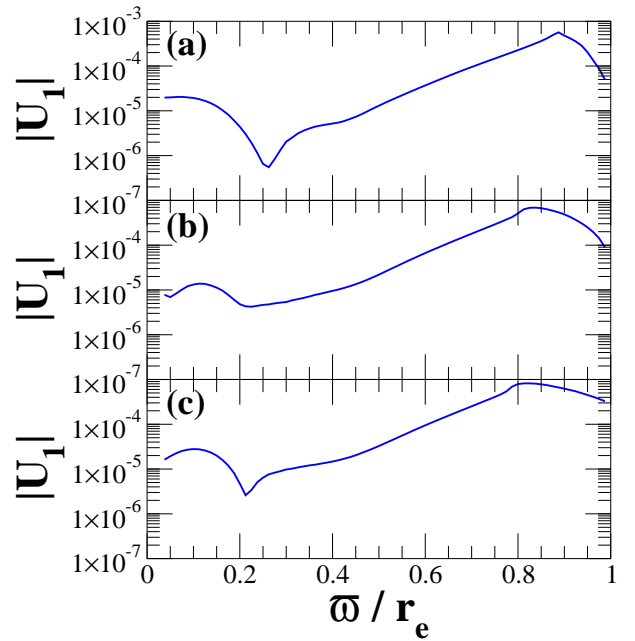}
\caption{
Same as Fig.~\ref{fig:figure3} but for $m=1$.  Note that $U_1$
represent the $m=1$ scalar potential density-weighted average  in the
equatorial plane.  The labels (a), (b), and (c) denote the evolution
time $t= 52.0~P_c$, $58.5~P_c$, and $64.9~P_c$, respectively.  We find
features similar to the cylindrical model (Fig.~\ref{fig:figure2}) in
which no nodes are present.
\label{fig:figure4}
}
\end{figure}

The real part of the eigenfrequency in the cylindrical model roughly
agrees with that of hydrodynamical simulation.  This situation is
improved to $\Re[\omega] = 0.352~\Omega_c$ and $\Im[\omega] =
0.0168~\Omega_c$ for model~I, $m=2$, $N=0$, when we take the $z$
structure of $\delta U_m$ and $\delta \Phi_m$ into account as
associated Legendre polynomial functions $P_m^m(z)$.  However, the
imaginary part of the eigenfrequency, which represents the growth
timescale of the instability, has few times difference with that of
hydrodynamic simulation.  This may be the fact that rotational
configuration of the stars is not fully taken into account in our
model.  The results of quasinormal modes of black holes suggest that
the imaginary part of eigenfrequency decreases as the black hole
rotation becomes fast for corotating modes \citep{Leaver85}.
Therefore, two-dimensional eigenvalue analysis of differentially
rotating stars is required for the full agreement of complex
eigenfrequency with corotation, but this, however, is out of the scope
of this paper.

We also compute the $m=1$ and $m=2$ diagnostic of scalar potential 
density-weighted average $U_{m}$ in the area of equatorial plane as 
\begin{equation*}
U_{m} = \frac{1}{U} \int_S ds~ue^{im\varphi}, 
\hspace{5mm}
U = \int_S ds~u, 
\end{equation*}
where $u\equiv H + \Phi=\varepsilon + P / \rho + \Phi$ and
$\varepsilon$ is the specific internal energy.  The diagnostics
$U_{1}$ and $U_{2}$ are regarded as ``eigenfunction'' when the system
possesses a dominant characteristic frequency and growth timescale, as
it is regarded as a ``single'' mode.  Our choice of low $T/W$
dynamically unstable star clearly meets this criteria.  We show our
eigenfunction in the equatorial plane computed from hydrodynamic
simulation in Figs.~\ref{fig:figure3} and \ref{fig:figure4}.  The
slope of the $m=2$ eigenfunction seems to have a sharp change in
model~I around the radius $\varpi / r_e \approx 0.45$ for all three
snapshots.  The $m=1$ eigenfunction seems to have a sharp change in
model~II around $\varpi / r_e \approx 0.20$ for all three snapshots.
Although hydrodynamics system may contain multiple ``modes'' in the
nonlinear regime, and they complicate the outcome, we find similar
behavior in the diagnostics as what is found in the eigenfunction by
linear analysis.

\section{Conclusions
\label{sec:Conclusions}}
We have investigated the unstable feature of low $T/W$ dynamical
instabilities in differentially rotating stars by means of normal mode
analysis in the equatorial plane.

We find unstable normal modes for low $T/W$ dynamically unstable stars
in the linear analysis.  Any additional modes to the well-known $f$
and $p$ modes  in the linear analysis are not found in our analysis,
and these modes become unstable when corotation radii exist inside the
stars.  The frequencies of the real part in the cylindrical model of
the linear analysis roughly agree with those of hydrodynamic
simulation.  The results confirm that our models are efficient for
finding low $T/W$ dynamically unstable stars.

We also find that the eigenfunction of the modes have a similar
behavior to the well-known $f$ and $p$ modes.  Once corotation exists
inside the stars, the perturbed enthalpy oscillates between corotation
and the surface radius.  Note that the perturbed enthalpy globally
oscillates between the center and the surface for the no-corotation
case.  This may indicate that the perturbed enthalpy is affected by
corotation singularity barrier, and therefore cannot cross the
corotation radius.  This feature requires reinterpretation of the
pulsation modes in rotating stars when corotation exists inside the
stars.

We have computed the linear analysis in the equatorial plane by
reducing the system to ordinary differential equations.  Our results
clearly show that rotational configuration of the star should be fully
taken into account for quantitative comparison to hydrodynamic
simulations.  To make a complete agreement between the linear analysis
and hydrodynamic simulation, a two-dimensional eigenmode analysis with
corotation is required, and it is a challenging task in this field.

\begin{acknowledgments}
It is our pleasure to thank Toni Font for valuable suggestions about
the shock capturing scheme to install in our hydrodynamic code.  This
work was supported in part by JSPS Grant-in-Aid for Young Scientists B
(Grant No.~201103201) and the Waseda University Grant for Special
Research Projects (Grant No.~2013A-6164).  Numerical computations were
performed on the Cray XC30 cluster in the center for Computational
Astrophysics, National Astronomical Observatory of Japan, and on the
cluster at Relativistic Astrophysics Group at Department of Science,
Waseda University.
\end{acknowledgments}


\end{document}